# Low-noise amplifier cryogenic testbed validation in a TaaS (Testing-as-a-Service) framework

Brandon Boiko[1], Eric J. Zhang[2], Doug Jorgesen[3], Sebastian Engelmann[2], Curtis Grosskopf[2] and Ryan Paske[2]

[1]FormFactor, Boulder, CO 80301 USA
[2]IBM Quantum, IBM T.J. Watson Research Center, Yorktown Heights, NY 10598 USA
[3]Marki Microwave, Morgan Hill, CA 95037 USA

Corresponding author: Brandon Boiko (email: brandon.boiko@formfactor.com).

This work was supported in part by the Quantum Economic Development Consortium (QED-C)

**ABSTRACT** As quantum computers based on superconducting qubit processors scale, cryogenic microwave components in the qubit control and readout chain must be appropriately tested and qualified to ensure consistent and high-fidelity quantum computation. However, the intersection of superconducting cryogenics and microwave electronics is a new domain with limited technical and commercial expertise. In this paper we validate a TaaS (testing-as-a-service) framework using an organizational workgroup model that consists of (1) a commercial Test House, (2) standard temperature Component Manufacturer, (3) Academic Partner, and (4) System Integrator to demonstrate a scalable model for the qualification of cryogenic microwave components. The goal of this model is to secure the supply chain and support the rapid growth of Quantum Computing (QC) technologies. The component test vehicle presented in this paper is a low-noise amplifier (LNA) which is a crucial component in the cryogenic chain to ensure adequate signal-to-noise of the qubit readout. We devise standard test metrics and protocols by which LNA performance is measured, including key parameters such as gain and flatness, reflection and isolation, operating bandwidth, and noise figure. We present details of the cryogenic testbed customized for LNA qualification, outline test methodologies, and present a suite of standard processes that are used to systematize data collation and reporting. The testbed is validated by reproducing parameters of a pre-characterized LNA. Its value is demonstrated by characterizing a proof-of-concept cryogenic LNA prototype. Finally, we describe the extension of our TaaS framework toward testing at scale for various active and passive cryogenic components used in QC.

## I. INTRODUCTION

Quantum Computing (QC) has rapidly emerged as an active technological growth sector driven by potential applications in security, applied chemistry, and finance [1-3]. Leading hardware modalities, particularly those based on cQED (circuit quantum-electrodynamics) architectures [4-6] have achieved processor scales beyond 100-qubits [7, 8] but require extremely low temperature ($\sim 10^{-2}$ K) environments to maintain their fragile quantum states. Each given qubit is associated with a cryogenic control and readout chain which is required for state preparation and measurement [4, 9, 10]. As the number of qubits scale to increase QC performance standards as measured by industry benchmarks (e.g., QV, *quantum volume* [11], or CLOPS, *circuit-layer operations per second* [12]), one of the limitations hindering a reliable QC supply chain is the ability for companies to provide microwave components that have been tested and qualified at cryogenic temperatures. However, the domains of cryogenics and microwave electronics are two highly specialized and rarely overlapping disciplines. The rapid increase in QC scaling [8, 13] drives the present need for dedicated test protocols and infrastructure for QC component testing. Numerous active (e.g., amplifiers, switches) and passive (e.g., circulators, isolators, cabling, and connectors) components will need to be qualified at cryogenic temperatures and microwave operating frequency bands.

A new product family typically requires several design-fabricate-test cycles to reach market readiness. Currently, Component Manufacturers are equipped to develop and qualify their components in the standard temperature range of -40 °C to 85 °C. The cost of establishing cryogenic testing capability and the naturally longer test cycles prevent Component Manufacturers from fully qualifying their products for the QC market. Thus, the ability to perform economical cryogenic component testing is critical as QC systems scale and are commercialized. Further, reliable cryogenic systems for testing at scale have only recently begun to mature [14, 15]. In the past, cryogenic refrigeration systems were heavily reliant on liquid helium which posed significant limitations due to operating costs, helium shortages, and specialized operator training. In recent years there has been a shift to *dry* systems [15-17], which operate as standalone cryostats without the need for cryogenic liquids, thereby significantly reducing strain on the helium supply chain. Future systems will continue to drive down operating costs by increasing utilization and reducing the burden of labor. For example, higher cooling power systems are being developed for larger loads [18] and faster cycling times. The use of load-lock chambers allows for cold-loading samples, thus eliminating the need to cycle the system [19, 20]. Software automation and mechanized sample exchange will further decrease labor time and operating expenses. Despite these advances, obtaining this equipment and skilled labor remains expensive and poses a significant barrier for developing and qualifying products for the QC market. Instead, centralized locations with cryogenic testing capabilities would reduce the overall labor and capital equipment burden on the industry. This equipment can be made available through widespread offering of cryogenic Testing-as-a-Service (TaaS).

Why doesn't this capability exist today? Two reasons stand out as most notable. First, there has not been significant business incentive from QC System Integrators (i.e., those entities driving the scale and development of QC technologies) for commercially available test infrastructure, due in large part to the research nature of QC [13, 21]. Second, investment into the unique skill set and infrastructure is too costly for most companies. For example, a Component Manufacturer with expertise in measurements within typical electronic component operating temperature ranges may not be willing to carry the burden of investment in cryogenic infrastructure and skilled labor in the absence of significant return on investment or adequate demand for TaaS.

Given the above, it is critical that TaaS be incentivized to foster a healthy QC market ecosystem. Failing to establish this capability will result in several key consequences that will impact the ability of the quantum industry to mature. For example, standard temperature Component Manufacturers will find it prohibitively difficult to enter the QC market given their lack of ability to develop and test components that meet the requirements of System Integrators. This will result in the component ecosystem being supported by only a small number of specialized Component Manufacturers. The absence of a competitive marketplace will stymie both the innovations and the competitive labor market required to rapidly advance and scale cryogenic component technologies for QC. Volume testing will remain difficult and burdensome as standard methodologies are not established to unify the market, and no industry standards are available to ensure accurate and repeatable product specifications. Without traceable and certified testing at scale, components will not have the necessary quality and reliability to perform as needed in a production environment.

A. CRYOGENIC TaaS WORKGROUP MODEL

Developing these TaaS capabilities requires a range of diverse skills from a variety of organizations and stakeholders to ensure robustness for widespread adoption. As a result, we set up our workgroup model (Fig. 1) with three goals in mind:

I. Develop and pilot the test framework for a Low-Noise Amplifier (LNA) for which extensive test knowledge already exists for device validation.

II. Make that framework available such that it could be used by others as a starting point to either test LNAs or apply our learning to testing other cryogenic commodities (e.g., isolators, circulators, etc.).

III. Generate interest in participation from potential TaaS parties to adopt these methodologies and generate new TaaS capabilities in the market.

The TaaS workgroup consists of four parties (as shown in Fig. 1), each of which serves a critical role in the success of the workgroup model. These parties and corresponding entities involved in the LNA TaaS demonstration in this paper are described below:

1. *Test House (FormFactor Inc.)* – provides technical expertise and the cryogenic test environment to facilitate component testing. The Test House will be the primary driver of component specifications and reliability testing and will conduct standardized and traceable cryogenic component qualification.

2. *Component Manufacturer (Marki Microwave Inc.)* – provides technical expertise in standard temperature microwave components. Develops the proof-of-concept cryogenic LNA and iterates on the design based on feedback from the Test House and Academic Partner.

3. *Academic Partner (Montana State University)* – collaborates with both the Test House and Component Manufacturer to provide technical

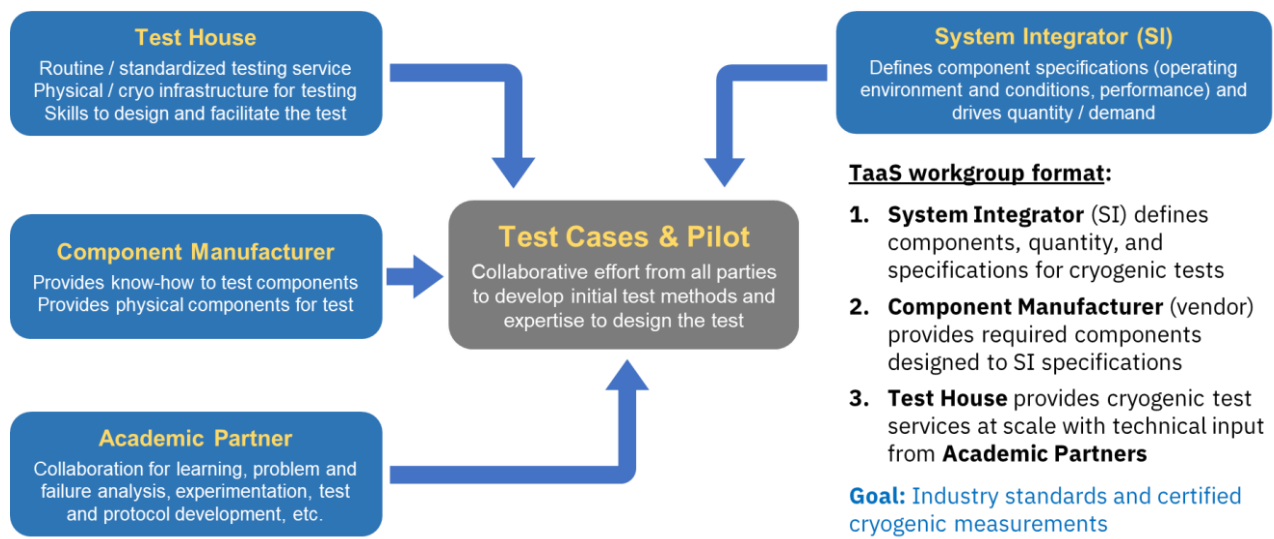

**Figure 1** The TaaS workgroup model for supply chain management of cryogenic microwave devices for QC. The four parties include: (1) a Test House with the cryogenic test infrastructure required to test the microwave components; (2) a Component Manufacturer (i.e., vendor) supplying the devices; (3) an Academic Partner with intimate knowledge of test methodologies and failure analyses; (4) a System Integrator with the ability to scale device demand and define the required specifications. A major goal of this framework is to define industry regulated standards for cryogenic component testing and specifications.

knowledge, failure analysis capabilities, and improved/optimized test methodologies.

4. *System Integrator (IBM Quantum)* – QC system provider, which drives the need and scaling roadmap for cryogenic components by defining required specifications (both performance and reliability), scaling demand, and directing the TaaS workgroup based on their needs and test requirements.

### B. BUILDING a TaaS ECOSYSTEM

The ultimate goal of this workgroup model is for it to grow into a self-sustaining ecosystem for scaling cryogenic test capabilities. The process should incentivize work on new technologies and provide avenues for commercialization. As such, the model must economically benefit all parties.

From the Test House perspective, TaaS is an opportunity to leverage their expertise and infrastructure for revenue generation and an opportunity to expand their application space and service portfolio. Initially, Test Houses will likely originate from cryogenic instrument manufacturers as they would have existing infrastructure and can use TaaS to advance their systems capabilities and development. As the application space for cryogenic components expands, smaller startups may take advantage of niche market segments. Furthermore, industry growth in QC and cryogenics will lead to the proliferation of TaaS users, increasing demand for more specialized markets to exist.

Component Manufacturers would gain access to otherwise prohibitively expensive capital equipment and technical expertise without being required to make a significant investment in cryogenic infrastructure and labor, thus lowering the risk in engaging in a new market. For Component Manufacturers with existing cryogenic capabilities, the TaaS model will allow them to supplement internal test capacity to support fluctuations in demand. Build cycles for QC systems often require all components simultaneously with large gaps between system builds making demand challenging to predict. The TaaS framework ensures robustness in the supply chain, providing a buffer against demand spikes without over-expanding operational capacity or inflating lead times.

The role of the Academic Partner is to provide guidance and expertise on testing new cryogenic technologies, and technical knowledge in failure analysis of cryogenic infrastructure and component devices. Academic Partners may also serve as objective entities that certify and validate Test Houses to ensure traceable measurements, thereby providing a guaranteed level of quality control in the cryogenic supply chain. They would also be able to offer specialized services to perform more detailed experimentation in cases where Test Houses may not have the required level of technical specialization or equipment. Academic Partners would therefore lead the research and development of advanced test capabilities to ensure that Component Manufacturers and Test Houses are aware of issues that may be difficult to identify. Support to the Academic Partner will be provided by all parties in exchange for key failure analysis, test protocol design, and new measurement techniques.

System Integrators would benefit from higher quality, more reliable, and lower cost components. A robust supply chain provided by the Test House, Component Manufacturer and Academic Partner will ensure easier access to cryogenic products and expertise, resulting in better components and reducing development cycle times. The outsourcing of testing capabilities in the TaaS workgroup allows the System Integrator to shift their focus towards QC system-level performance and reliability (as they are ultimately responsible for providing a working product to the end user) instead of being limited by failures and errors resulting from individual components. Currently, much of the component-level testing burden is on the System Integrator, which limits the economic scale at which large commercial systems might be deployed, particularly as QC systems scale beyond $10^3$ qubits. Cryogenic TaaS would shift much of the testing workload back to the market where it can be scaled more efficiently given higher levels of competition, leading to faster development and more reliable components at lower costs.

The present challenge facing cryogenic TaaS is in establishing a self-sufficient market. The benefits and roles of each TaaS workgroup member will not materialize until a critical market size is reached. Both users and providers of cryogenic test services must be incentivized to explore the component application and testing space. Existing resources must be leveraged to develop new cryogenic technologies and test methodologies, with the reward being first mover advantage to a rapidly growing market.

### C. TEST CASE: THE LOW-NOISE AMPLIFIER (LNA)

The pilot test case for our TaaS workgroup is the low-noise amplifier (LNA). This choice was made because LNA characterization is a difficult measurement that serves as a good representation of the specialized services required for testing cryogenic electronics for QC. In addition, there has been extensive device study [22, 23] and documentation into performing this measurement accurately [24, 25], providing us with the knowledge to quickly set up the test and measurement equipment (TME) and fulfill our goal of demonstrating the TaaS workgroup for this use case. LNAs are also highly relevant to the scaling of QC systems, given that the number of required LNAs increases proportionally with the number of qubits [10]. Cryogenic LNAs are available from a small number of commercial vendors, however at current prices and levels of LNA reliability, serious financial and technical feasibility concerns emerge as QC systems continue their rapid scaling.

As described previously (Section I-A), our test framework is based around four principal parties. The untested proof-of-concept device was provided by the Component Manufacturer (Marki Microwave Inc.), along with preliminary design operating characteristics (e.g., design frequency operating range). The devices were received by the Test House (FormFactor Inc.) whose responsibilities included providing and developing the cryogenic TME. Finally, with guidance and collaboration from both the System Integrator (IBM Quantum) and Academic Partner (Montana State University), the test methodology and measurement parameters (Section III) were defined. Our hope is that in demonstrating this first functional TaaS workgroup and the necessary TME and measurements, our model may serve as a blueprint that will be extended towards testing other RF and microwave components in the QC cryogenic chain.

## II. CRYOGENIC SETUP AND TEST PROTOCOL

The measurements in this paper were performed at the Test House (FormFactor), using a two-phase test protocol developed and agreed upon by all TaaS members. The purpose of our test scheme was to ensure the cryogenic test and measurement equipment (TME) was operating to documented specifications to ensure comparable test results.

### A. CRYOGENIC TEST AND MEASUREMENT SETUP

The measurements were conducted in a HPD Model 106 ADR (adiabatic demagnetization refrigerator) cryostat [26]. However, any 4 K capable system can be used. The radiofrequency (RF) signals are input through the top of the cryostat through a series of rectangular flanges and feedthroughs in the vacuum and temperature stage plates. Each end of the coaxial cabling, running between the 50 K and 4 K plates, is terminated with a 0 dB attenuator for thermalization. A sample platform to mount the LNAs and switch electronics is mounted to the underside of the 4 K plate.

Fig. 2 shows the inside of the cryostat with cabling running between the temperature stages. This cabling is semi-rigid beryllium copper (BeCu) for thermal isolation. Not shown are the radiation shields at both 4 K and 50 K. Inside the 4 K measurement space we use flexible .047" copper coaxial cables. High resistance cabling is not required since both ends are terminated at 4 K.

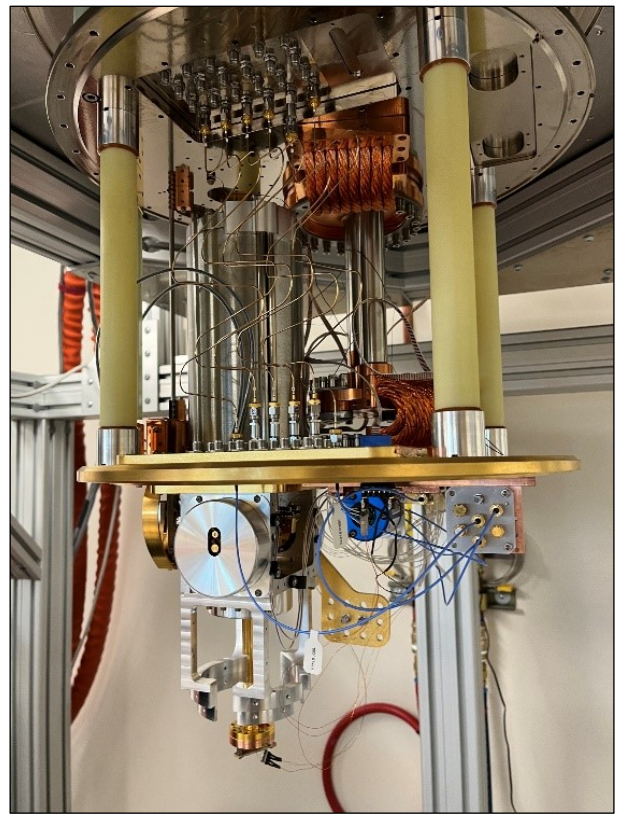

**Figure 2** Overview of the FormFactor HPD Model 106 ADR cryostat measurement system [26]. Two coaxial cables are required for the input and output of the system and are connected to either the VNA for S-parameter measurement or the SA for noise analysis. The system is equipped with 6 semi-rigid BeCu coaxial lines for good thermal isolation. Attenuators are used at each end of the cabling for thermalization. At the 4 K stage, two SP6T switches are used to perform calibrations.

The sample platform (Fig. 3) consists of a 101-copper adapter plate with brackets for mounting two Radiall SP6T cryogenic switches. Each switch is populated with cryogenic calibration standards (XMA Corporation). A THRU-REFLECT-LINE (TRL) set is used to perform the full TRL calibration [27, 28]. The device under test (DUT) is placed between the two switches on an open port. A calibrated Cernox thermometer (model: CX-1050-CD-1.4L) is installed near the LNA to accurately read out the DUT temperature. A complete list of all components and equipment used in this measurement is provided in Appendix A.

Two measurement configurations are required to gather the desired device parameters. The first configuration (Fig. 4a) shows Port 1 and Port 2 of the VNA (Keysight P5025B) connected directly to the input and output ports of the cryogenic system for S-parameter measurement. The second configuration (Fig. 4b) depicts a SA (N9010B EXA) setup for noise figure analysis, where a calibrated noise source (Noisecom NC3609) is connected to the system input, while the output readout connects to the SA. A 30 dB attenuator is inserted before the DUT to provide a cold reference noise source. Another Cernox (model: CX-1050-CO-HT-1.4L) was potted to the attenuator, Fig. 5, using black stycast epoxy to provide a more accurate measurement of the cold reference temperature.

The cryogenic setup depicted in Fig. 2 and 3 is capable of testing single LNAs. However, our cryogenic TME is easily extensible to testing multiple LNAs by expanding the number of switch units to sequentially cycle through an LNA array. Finally, we note that although the cryogenic TME is depicted here for LNA characterization, the general best-practice principles of calibration, measurement and thermalization in these cryogenic systems hold regardless of the nature of the DUT under study.

### B. LNA TEST PROTOCOL

Our cryogenic testing is centered around two phases using two respective LNA devices which we identify as LNA-C (Control) and LNA-T (Test). The first device, LNA-C (Low Noise Factory, model LNF-LNC4_8C, S/N: 3225H) [29], was provided with known specifications and used to qualify our cryogenic TME and methodology (described in Sections II and III). By measurement and comparison of LNA-C against factory specifications, we could determine whether our TME outcomes yielded areas of discrepancy that might indicate the need to refine our cryogenic setup or test protocols.

The second device, LNA-T (Marki Microwave Inc.), was created as a proof-of-concept to determine whether an operationally functional LNA could be created by a high frequency electronics company with no prior expertise in cryogenic design or manufacturing techniques. The design for this amplifier was based on a specification provided by the system integrator and was designed, fabricated, and tested using only techniques, software, and equipment that were common with the manufacturer's standard temperature products. Only small modifications were made to eliminate the use of sub-components known to fail at cryogenic temperatures. The cryogenic performance of this amplifier and the processes used to produce it were unknown prior to testing. The test sequence phases and outcomes are depicted in Fig. 6.

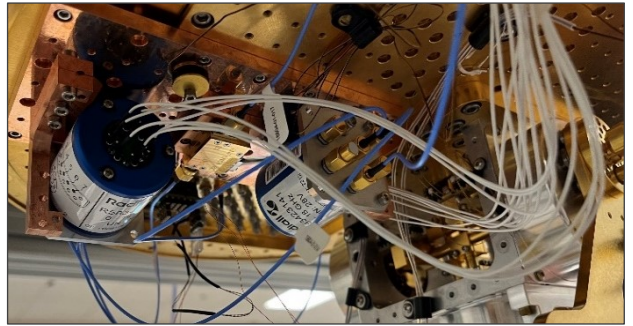

**Figure 3** Sample platform mounted to the underside of the 4 K stage plate. Two SP6T switches mounted to either side of the DUT (the LNA shown is an LNF-LNC4_8C [29]). A calibrated Cernox thermometer is also attached next to the LNA for accurate device temperature measurement.

In both Phase 1 and 2 of testing, a standard set of parameters to benchmark LNA-C and LNA-T were measured using the test infrastructure and methodology outlined in Section III. During each measurement phase, the qualification device (LNA-C, Phase 1) or test device (LNA-T, Phase 2) were identically mounted to the cryogenic test fixture to ensure minimal differences from cabling, connectors, and attenuators. Systematic calibrations were performed on both the P5025B VNA and N9010b EXA analyzers (used respectively for S-parameter and Y-factor noise measurements).

### III. TEST METHODOLOGY

The basic performance parameters for an RF device are response, linearity, and noise. Characterization techniques at room temperature involve proper calibration of the test fixture (i.e., removal of the impact of the fixture on DUT measurements), which is achieved with standards and modules that are measured in place of the DUT [30]. The RF response is readily measured by either scalar or vector network analyzers. Noise measurements are more involved, and the appropriate choice of methodology depends on the desired outcome and on the TME setup capabilities. These outcomes and corresponding methods are considered below.

When measuring devices with low noise figures, two main techniques exist: the Y-factor method and the Gain Method [31, 32]. The Y-factor measurement involves measuring the output noise power at two distinct input noise temperatures. The absolute power accuracy of the measurement equipment is less of a concern due to the use of relative hot and cold measurements [33]. In contrast, the Gain Method only measures the cold noise after the device gain has been very accurately characterized and requires only a single noise power measurement. Scalar noise figure measurements involve all cases where only 50 Ω impedance networks are considered. However, noise figure is also dependent on the reflection characteristics at the input of the DUT, also known as the vector noise parameter measurement. To paint a full picture of device noise, the DUT must be measured with varying input impedances [24]. This is achieved by manually replacing the input load in the Gain Method or using a noise source tuner in the Y-factor method.

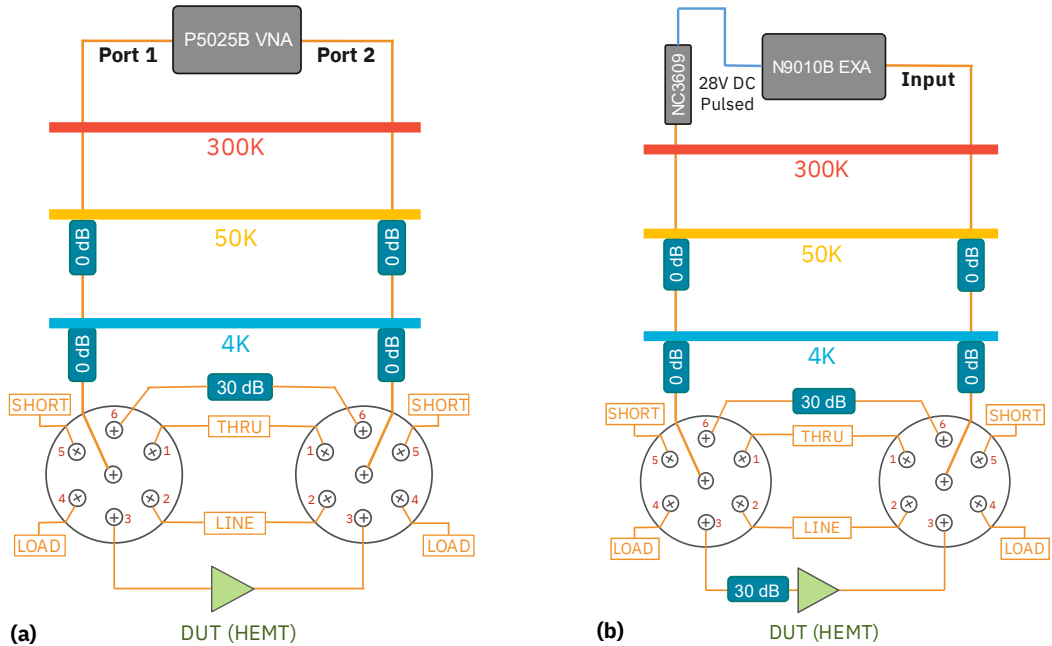

**Figure 4** TME configurations for S-parameter and noise figure measurements. (a) VNA setup to measure RF characteristics of the DUT. (b) SA set up to measure noise figure using the Y-factor method. In both cases, 0 dB attenuators are used for noise thermalization across temperature stages.

At cryogenic temperatures the method is the same, but many calibration modules are not designed to function at cryogenic temperatures and the DUT is not easily accessed to swap out calibration standards. Significant work has been presented in literature to overcome these limitations. For example, calibration has been achieved for both one port and two port configurations by using cryogenic switches to multiplex the input and output channels [28, 34]. Cryogenic noise figure is measured in the same way as the room temperature method but requires a controlled input noise source, which is either a cold attenuator or a cold load, used respectively in either the Y-factor or Gain Method [35]. Additionally, a matched variable temperature noise source has been demonstrated to function between 100 mK and 5 K for measuring quantum limited amplifiers with noise temperature down to 680 mK [25]. Instead of using an external noise source this device is mounted internally and the temperature is well controlled using heaters and calibrated sensors. Vector noise measurements can also be implemented in this configuration. Finally, we note that cryogenic impedance generators are commercially available [36] and may be used in cases where scalar noise measurements are insufficient.

For this paper we selected the cold attenuator Y-factor method, given the relative simplicity of cryogenic wiring required for extracting the noise figure. While the variable temperature noise source in the Gain Method would yield more accurate results, it would require additional switches to swap the input load and was deemed unnecessary for the scope of this paper. The Y-factor method provides adequate accuracy for LNA noise characterization, and scalar noise measurements are sufficient for measuring LNA-C and LNA-T specifications for the purpose of our TaaS demonstration.

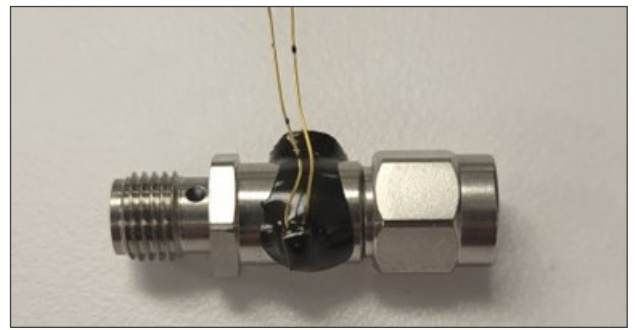

**Figure 5** 30 dB attenuator with a Cernox temperature sensor potted to the outer shell using black stycast epoxy. This sensor provides a more accurate temperature measurement for the cold reference used in the noise figure measurement.

### A. CALIBRATION

Accurate measurements with both the VNA and the SA require the equipment to be calibrated to the test fixture. This is accomplished by measuring standards with well-known responses and removing systematic errors from subsequent measurements. For cryogenic applications, these standards must have well-known characteristics at the desired measurement temperature.

The VNA is used to measure basic RF parameters with the configuration in Fig. 4a. Loss and reflection characteristics of the test fixture are removed from the measurement during calibration. This effectively moves the plane of reference for the RF signal to the input and output of the DUT.

This in-situ calibration is performed by connecting the VNA port to each of the cryogenic standards: THRU, REFLECT (SHORT), and LINE (i.e., the TRL calibration method). Using built-in Keysight software, an ideal TRL basic calibration is measured and applied as a correction. The guided calibration requests each standard prior to taking the measurement. The SP6T is manually actuated by applying 28V DC to the relevant control terminals, but this process may be easily automated in the future to expedite the switching process. A successful calibration can be verified by reconnecting the THRU standard and checking the baseline correction is as expected. After TRL calibration, the insertion loss of the THRU should be 0 dB, corresponding to a null baseline.

Noise figure is measured using the SA in the configuration shown in Fig. 4b. Noise contributions from the system must be removed from the measurement, otherwise the TME will attribute it to the DUT. To achieve this, the size of the cold reference attenuator is carefully selected to minimize the impact of the system noise. Additionally, cable noise can be directly measured and de-embedded from the measurement.

Excess noise ratio (ENR) data of the calibrated noise source is first loaded into the SA. An initial standard calibration is performed with the noise source connected directly to the input of the SA. This is identical to the room temperature calibration process. A cryogenic calibration is described further in the following section.

### B. NOISE TEMPERATURE

The method used for measuring noise temperature of a device is the Y-factor method. Most spectrum analyzers and noise figure analyzers have automated this measurement making it one of the most common methods for analyzing noise. The challenge is obtaining a good calibration. Keysight's Noise Figure Measurement Application allows for automated measurement using the Y-factor method as described by the following equations.

$$Y = \frac{N_{out,hot}}{N_{out,cold}}, \quad (1)$$

$$T_{DUT} = \frac{T_{hot} - Y \cdot T_{cold}}{Y - 1} \quad (2)$$

The output noise in Eq. (1) is measured at the noise analyzer receiver and is affected by the loss of the system cables. Similarly, the hot and cold temperatures produced by the ENR diode are affected by cable loss and the cold reference attenuator. The noise contributed by the cable and attenuator is removed from the measurement by applying a loss table in the software.

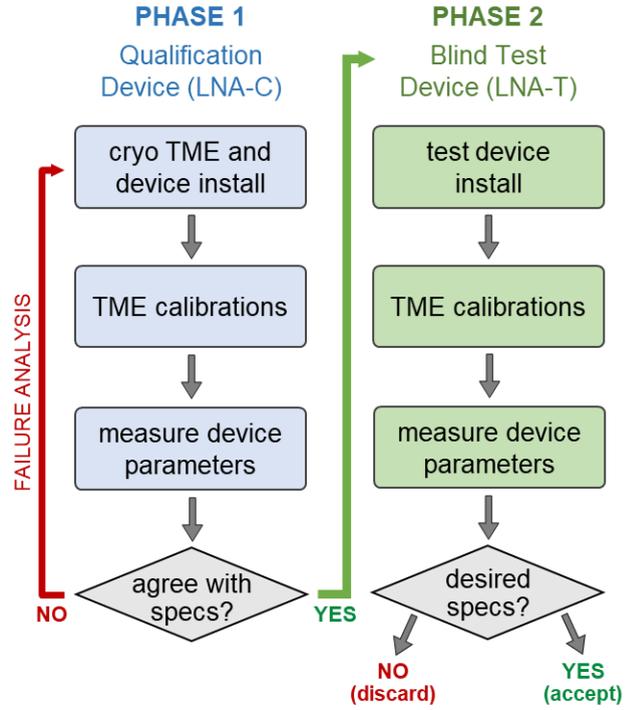

**Figure 6** TaaS cryogenic qualification and test sequence. In Phase 1, a qualification device (LNA-C) with known device parameters is installed into the cryogenic test and measurement equipment (TME) setup and measured to confirm agreement against known specifications (or if failure analysis is required). If the Phase 1 qualification is successful, the cryogenic TME may be deemed appropriately qualified and can proceed to Phase 2 where measurement of Manufacturer devices with unknown parameters are performed. If test devices do not conform to desired specifications (defined by the System Integrator and Manufacturer), they may be discarded or be subject to further failure analysis.

The loss table is constructed from an insertion loss measurement of the system without the DUT. This curve is split between the *before* and *after* DUT loss tables. The attenuator is measured during a previous cooldown and added to the *before* DUT loss table. Keysight's software corrects the measurement by removing the noise figure contribution from this loss at a specified input temperature, $T_{Loss}$. The temperature from the potted calibrated Cernox thermometer is used to determine the cold attenuator temperature $T_A$ and calculate the effective input noise temperature $T_{in}$ to the LNA. $T_{in}$ is a combination of the noise produced by the ENR diode $T_S$, cable noise $T_{cable}$, and the attenuator noise $T_A$

$$T_{in} = \frac{1}{L_A L_{cable}} T_S + \frac{1}{L_A}\left(1 - \frac{1}{L_{cable}}\right) T_{cable} + \left(1 - \frac{1}{L_A}\right) T_A. \quad (3)$$

Here $L_{cable}$ and $T_{cable}$ are continuously changing along the length of the cable making the second term in Eq. (3) challenging to model. Furthermore, the noise analyzer software allows for only a single temperature to apply to the loss compensation. A simplified model lumps both cable and attenuator loss into the single temperature $T_{Loss}$. Where $T_{Loss}$ is the effective noise temperature referenced to the output of the cable and attenuator system.

$$T_{in} \approx \frac{1}{L_A L_{cable}} T_s + \left(1 - \frac{1}{L_A L_{cable}}\right) T_{Loss} . \quad (4)$$

A model was created by dividing the cable into sections with measured end point temperatures. This model consists of a room temperature Cu cable (RRR = 100), a 50 K BeCu cable, and a 4 K BeCu cable for both input and output. Each section of the cable is 1 m, 0.5 m, and 0.5 m respectively. The cables were broken into 1000 discrete elements, for a total of 6000 elements along the length of the cable. Each element is treated as a separate attenuator as governed by the third term in Eq. (3). Temperature dependent material properties were used to calculate the gradient along the length of the cable [37-39]. Fig. 7a shows the calculated temperature gradient from the SA to the 4 K Plate. The effective noise temperature of the cable can be integrated using the temperature and loss of each element. Alternatively, the effective noise temperature can be calculated with a lumped cable temperature using the equation below.

$$T_{eff} = (L_{cable} - 1) T_{cable} , \quad (6)$$

This curve was fitted to the integrated model by adjusting $T_{cable}$ to minimize the error, resulting in an average temperature of 210 K. $T_{Loss}$ in Eq. (4) can be calculated using this temperature in the following equation.

$$T_{Loss} = \frac{(L_{cable} - 1) T_{cable} + L_{cable}(L_A - 1) T_A}{L_A L_{cable} - 1} \quad (7)$$

Both the integrated and lumped temperature curves are superimposed on top of data from a SA measuring the noise of a THRU standard, Fig. 7b. The discrepancy between these two curves is due to the small frequency dependance of $T_{cable}$. Both curves seem to follow the trend of the measured data, though they appear slightly lower at the higher frequencies. The difference between the model and the data could be caused by model error due to underestimating the average temperature of the loss. Alternatively, some parts of the cable may have been warm when the measurement was taken, resulting in a higher noise temperature. This can be verified in the future by allowing the system to rest at base temperature for longer before measuring.

Two things must be considered when selecting an attenuator for this measurement. Too much attenuation may reduce the source power below the measurement capabilities of the SA. Alternatively, too little attenuation will lead to a higher uncertainty due to cable noise.

## IV. RESULTS
Following the protocol in Section II-B, our results have been summarized in two sections below, respectively describing (A) the qualification of our cryogenic system using LNA-C, by comparison against manufacturer specifications (as provided by Low Noise Factory for this specific device), and

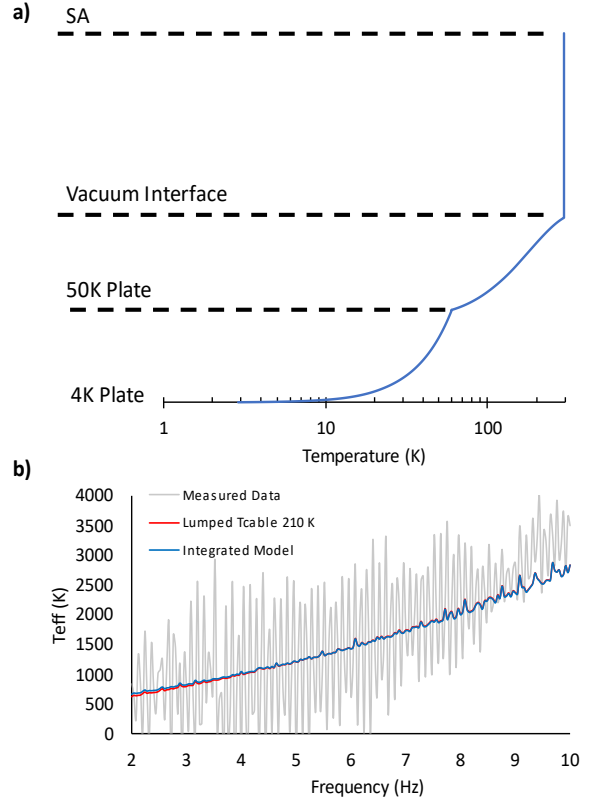

**Figure 7 (a)** Model of the temperature gradient along the length of the cable. Measurements from an ambient thermometer and diode sensors formed the basis for the model. **(b)** Effective noise temperature referenced to the input of the cable. Integrated model using temperature dependent properties (blue). Lumped model at 210 K (red). Measured Teff of a THRU Standard (grey).

(B) the generation of a specifications report for LNA-T (i.e., the test device provided by the Component Manufacturer, Marki Microwave Inc.). All test methodology and parameters follow the discussion outlined in Section III.

### A. Phase 1 (LNA-C): Cryogenic TME Qualification
All measurements were performed on LNA-C with the bias conditions at 4 K set to $V_D = .7\ V$, $I_D = 15\ mA$, and $V_G = -1.0\ V$. Fig. 8 shows the $S_{21}$ gain spectrum measured at 2.88 K for LNA-C, with the blue and orange curve showing uncalibrated and TRL calibrated gain spectra respectively. Cable loss is the dominating contributor to the uncalibrated offset as the THRU shows a good match at 1.2 VSWR. The TRL calibrated measurements (Fig. 8, orange curve) were an average of 2.0 dB below the manufacturers data, this is attributed to a different bias condition as indicated by a 40 mV difference in gate voltage. Based on the calibrated gain spectrum, we determined a gain flatness of $GF = 1.0\ dB$ over the 4-8 GHz operating range which matches the manufacturer provided data.

HEMT LNAs exhibit nonlinear behavior with output gain saturation at high input powers. This input power-dependence is quantified by the *input 1 dB gain compression* (*IP1dB*), which is the power above which the gain spectrum is reduced by 1 dB and whose value determines the input dynamic range of the amplifier. *Output 1 dB gain compression* is often reported as well as a good indicator of the amplifier efficiency. Fig. 9a shows a power sweep from -80 dBm to -20 dBm taken at 6.0 GHz plotted against the response. A linear interpolation of the region from -80 dBm to -60 dBm is projected across the sweep range. The point at which the output power deviates from this line is the measured P1dB point. Both output and input P1dB are denoted by the dotted grey lines. Fig. 9b shows the dependence of the OP1dB compression on input frequency.

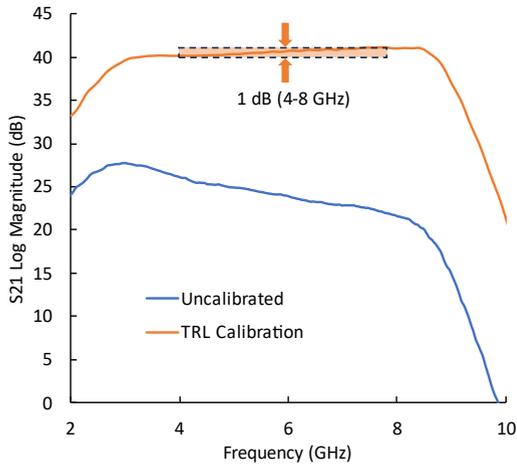

**Figure 8** $S_{21}$ gain spectrum at 4 K of LNA-C mounted in the cryostat. The blue and orange curves show the uncalibrated and TRL calibrated measurements. The substantial discrepancy is due to the large cable loss and demonstrates the necessity of adequate calibration prior to measuring device S-parameters. Based on the TRL calibrated gain, we find a gain flatness of 1.0 dB

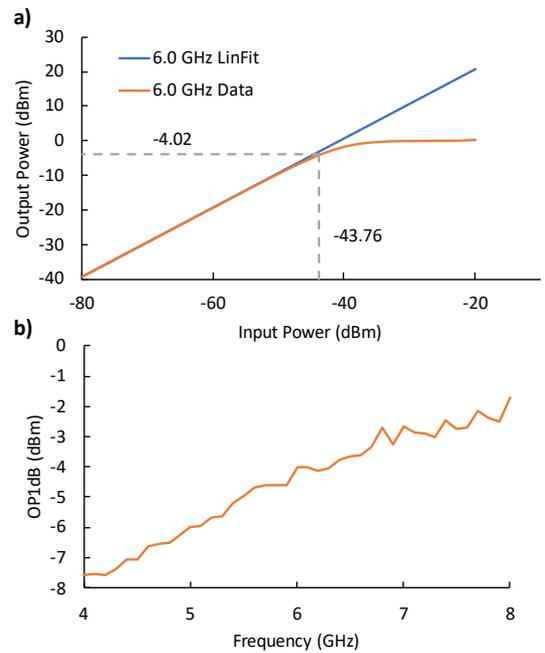

**Figure 9** Gain compression (P1dB) measurement of LNA-C. (a) A input power sweep plotted against the measured gain response (orange curve) a linear interpolation of the region -80 dBm to -60 dBm projected across the entire sweep (blue curve). P1dB compression points (grey dotted lines) (b) Extracting OP1dB compression for frequencies spanning from 4-8 GHz (the design operating range of LNA-C).

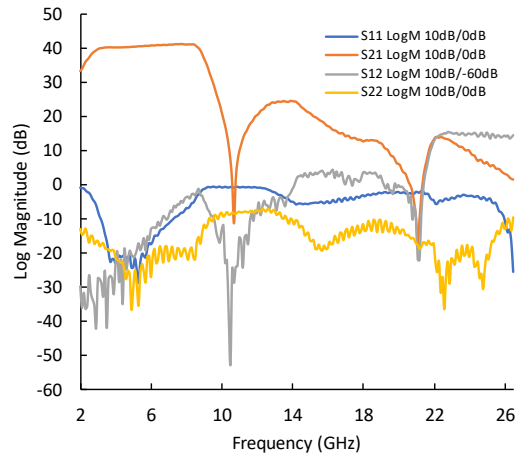

**Figure 10** TRL calibrated S-parameter measurement for LNA-C. Log magnitude measured from 2 GHz to 26.5 GHz. S12 (grey curve) is depicted with a 60 dB offset.

In the present case for LNA-C, gain compression analysis was not provided in the device specifications sheet; instead, the manufacturer (LNF) provided a general OP1dB target in the range of -10 dBm, which is in coarse agreement with our values. We anecdotally note that typical reported values are conservatively estimated which results in an effectively higher output gain compression value under practical use.

In addition to the commonly reported $S_{21}$ gain spectrum, other S-parameters are also important metrics to quantify isolation behavior and input/output port reflections. With regards to the latter, the $S_{11}$ and $S_{22}$ parameters were measured by the VNA to be below -10 dB within the span of 3.2 GHz to 7.3 GHz, in agreement with manufacturer's data. $S_{11}$ and $S_{22}$ are measures of input and output return loss respectively, with lower values indicating less reflections, which is desirable to mitigate power feedback/dissipation within the cryostat.

Our $S_{12}$ measurement is found to yield < -60 dB of reverse isolation, indicating good signal isolation and minimal leakage. We note in passing that in our present measurement configuration, a number of repeatable standing wave patterns are observable in our S-parameter measurements, the likely result of long etalon paths in the cryogenic TME chain. In the future, such etalon artifacts can be removed by use of appropriate attenuators and improved impedance matching to avoid spurious back-reflections.

Noise temperature was measured and compared against data provided by the manufacturer, Fig. 11. Measurements were performed using the temperatures from the two separate Cernox sensors on the test fixture. With the amplifier turned off, both sensors measure a base temperature of 2.74 K. With the amplifier on, the sensor mounted near the base of the LNA measured a $T_A$ of 2.88 K, and the sensor that was potted to the attenuator measured a $T_A$ of 3.2 K. Consequently, the noise temperature is offset by an average value of 380 mK within 4-8 GHz. We note the similarity with the 320 mK offset of the physical temperature. Consistent with the relation of $T_{DUT}$ with $T_{cold}$ for sufficiently high Y factor in Eq. (2). This highlights the importance of accurately measuring the attenuator temperature as any uncertainty in this temperature directly translates to uncertainty in the noise temperature. An analysis was performed by compounding the measurement uncertainties of each parameter, and are tabulated in Table I. We determine a combined uncertainty for the DUT noise temperature of 150 mK which is indicated in Fig. 11 by the black error bars. We also note that the measurements deviate from the manufacturer's data at high frequencies beyond 6.0 GHz. This could be caused by poor impedance matching at those frequencies.

Calibration repeatability is important to characterize the stability of the measurement equipment and consistency of the method. A fresh calibration was performed on 9 separate measurements over 8 hours. The 95% confidence interval, Fig. 12, is plotted against frequency. The results show good consistency within the operating bandwidth of the device.

The calibrations and measurements described above serve to both qualify our cryogenic TME system and verify the vendor specifications of LNA-C. In general, qualification of new cryogenic and RF test infrastructure will involve similar benchmarking using a test device with well-known performance specifications, such that agreement (or discrepancies) may be used to validate or improve the cryogenic measurement system.

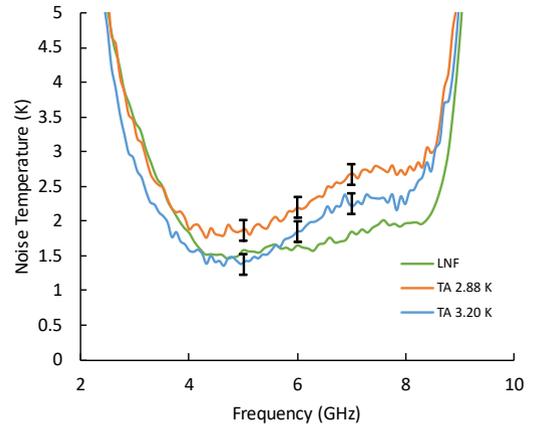

**Figure 11** Noise temperature measurements of two different attenuator temperatures $T_A$ based on cernox measurements, compared with data provided by the manufacturer (green). $T_A$=2.88 K (orange) uses the temperature measurement of the cernox adjacent to the LNA. $T_A$ = 3.20 K (blue) is the temperature measured by the potted cernox on the 30 dB attenuator. Black error bars showing +/- 150 mK uncertainty.

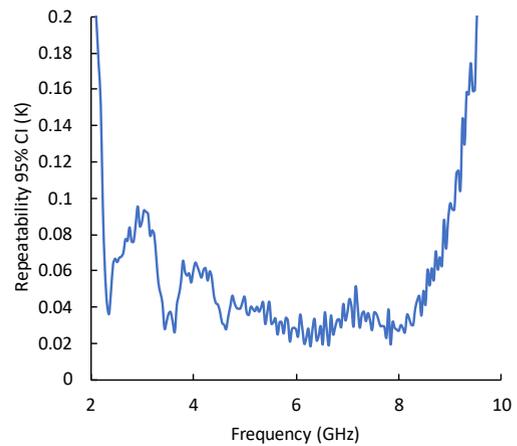

**Figure 12** Calibration repeatability determined from performing the same calibration over 9 separate measurements. Two standard deviations plotted as a function of frequency show <100 mK across the measurement bandwidth. This is consistent with the 150 mK uncertainty calculated in Table I.

TABLE I
MEASUREMENT UNCERTAINTIES

| Parameter | Uncertainty |
|---|---|
| $G_{DUT}$ | .033 dB[a] |
| $L_{Cable}$ | .033 dB[a] |
| $L_A$ | .033 dB[a] |
| $T_{Cable}$ | 32 K[b] |
| $T_A$ | 5.0 mK[c] |
| ENR | .18 dB |
| $T_{eff}$ | 12 K[d] |
| $T_{DUT}$ | 150 mK |

[a] The calibrated measurement uncertainty of the VNA
[b] Average cable temperature uncertainty based on the measured data compared to the model in Section III
[c] Accuracy of the calibrated Cernox RTD
[d] Uncertainty in the SA noise temperature measurement, dominated by the ENR uncertainty

### B. Phase 2 (LNA-T): Device Specifications Testing

Given the qualification of our cryogenic TME setup from Phase I (Section IV-A), LNA-T was provided by the Component Manufacturer (Marki Microwave Inc.) with only basic device design parameters (e.g., operating frequency range and 4 K LNA bias conditions) to avoid biasing our TME results. LNA-T was tested by simply replacing LNA-C in the same cryogenic chain to maintain measurement consistency.

The proof-of-concept LNA-T was tested using the bias conditions $V_D = 0.5\ V$, $I_D = 7.8\ mA$, $V_g = 0\ V$. It showed functionality at cryogenic temperatures. Future design iterations will further hone the component specifications to match the system integrator requirements. An example of the $S_{21}$ gain spectrum for LNA-T is shown in Fig. 13. The design frequency range (i.e., operating bandwidth) for the device was 6-9 GHz, but for consistency of testing we used the same broader measurement range from 2-10 GHz as was done for LNA-T. Based on the 6-9 GHz operating range, we determined peak gain of 35.7 dB, with a mid-range (7.5 GHz) gain of 33.4 dB and a gain flatness of 3.5 dB. Room-temperature measurements (296 K) were also performed as an additional data set (Fig. 13, orange curve). A TRL calibration was performed at 4K whereas an e-calibration was performed at room temperature.

Gain compression measurements were performed by sweeping the input power to LNA-T from -80 dBm to -20 dBm and extracting the 1 dB compression point, in the same manner as that performed for LNA-C. The results of these measurements and the extraction of OP1dB is shown in Fig. 14. The amplifier showed good linearity up to 1 dBm with the OP1dB point reaching nearly 2 dBm at higher frequencies.

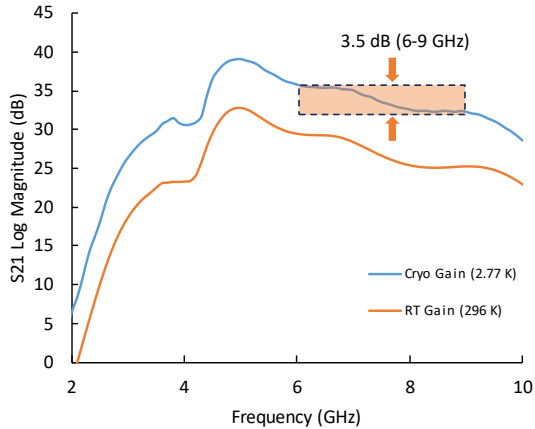

**Figure 13** $S_{21}$ gain spectrum of LNA-T (Marki Microwave Inc., S/N: 07711) at 4 K (blue curve). The intended device operating range was 6-9 GHz, with a gain of 38 dB. Measured gain flatness on the device was 3.5 dB (from 6-9 GHz operating BW). For comparison, the device was characterized using the same drain voltage and current at room temperature (296 K).

Calibrated S-parameters were also measured for this device, provided in Fig. 15. The device demonstrates good isolation, S12 <-60 dB, and good match, S11/S22 <-10 dB, within the operating bandwidth. Data was also provided for higher frequencies showing isolation is lost above 10 GHz.

Fig. 16 shows the noise temperature measurement of LNA-T with a value varying from 6 K up to 8 K within the 6-9 GHz operating bandwidth. This performance is in line with the manufacturer's designed specification of 6 K.

## V. DISCUSSION

The methods used for LNA characterization were selected based on existing technology and the simplicity of the measurement. In the case of our noise measurements, noise figure modules on many SAs and VNAs are readily available for the cold attenuator Y-factor method. However, other methods may provide better paths for scalability, or generate more comprehensive results. For example, the Gain Method [35] would require additional switches and software development to extract the noise figure but would eliminate the need for a second setup. The tuner method [24] would allow for a vector noise parameter measurement, providing data with different input impedances from which a model could be generated and offers a more complete picture of the noise figure.

Our current system has the capacity to test a single LNA and requires different configurations for S-parameter and noise figure measurements. To truly be useful and scalable in a production environment, parallel testing is needed to test all measurement parameters on multiple devices. A system with additional switches would allow multiplexing of the calibrated channels, which may be feasible for low to medium volume

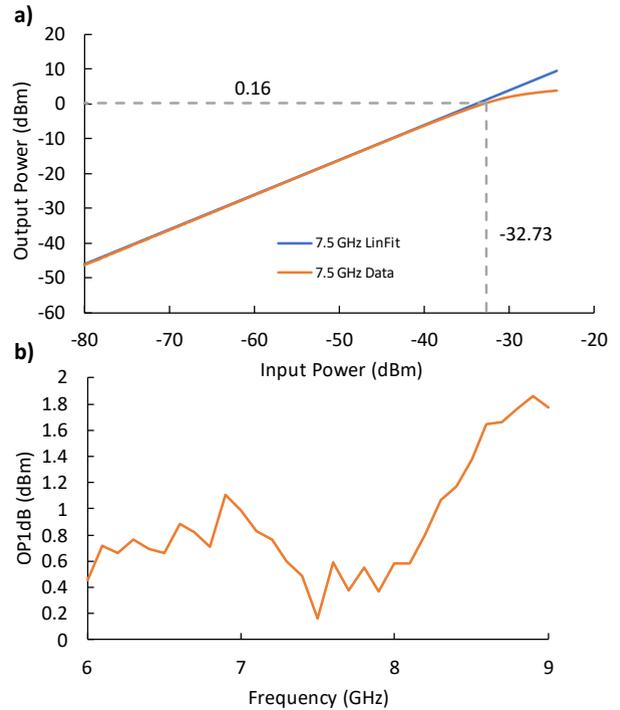

**Figure 14** Gain compression (P1dB) measurement of LNA-T. (a) An input power sweep plotted against the measured gain response (orange curve) a linear interpolation of the region -80 dBm to -60 dBm projected across the entire sweep (blue curve). P1dB compression points (grey dotted lines) (b) Extracting OP1dB compression for frequencies spanning from 6-9 GHz (the design operating range of LNA-T).

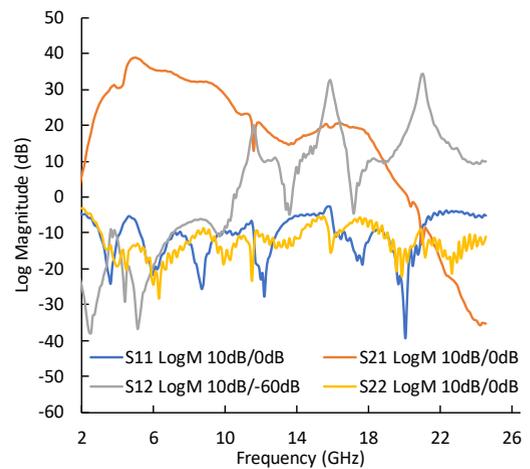

**Figure 15** TRL calibrated S-parameter measurement for LNA-T. Log magnitude measured from 2 GHz to 26.5 GHz. S12 (grey curve) is depicted with a 60 dB offset.

TaaS (10 – 100 devices). As the market develops and demand grows, volume testing will become a necessity [13].

Standardized device specification reporting will be required as TaaS scales, to maintain consistency across Test Houses. System Integrators and Component Manufacturers will require test results to be certified and traceable as they will rely heavily on the feedback of test reporting. In the context of TaaS, the Academic Partners, System Integrators and external impartial entities may be involved in this certification process by standardizing test methodologies and protocols (e.g., defining calibration standards), as well as ensuring any software or computational tools may be sourced by the community to avoid reporting discrepancies. Standardized benchmark devices may also be distributed to ensure all Test Houses are qualified using identical test components.

Additionally, a reliability test methodology needs to be developed to understand and predict failure modes over the lifetime of cryogenic components. This is a more challenging problem as thermal cycling in existing cryostats is resource intensive and requires expensive and slow cooldown/warmup cycles. Standard thermal shock methods (e.g., immersion in liquid nitrogen) do not apply as the stresses would be less extreme under normal conditions. Test qualification of any reliability methodology may also pose a challenge given that failure modes will likely be unknown and may materialize under conditions that are difficult to replicate. Significant innovations in reliability test methodology will be required to ensure that any tests correctly capture the life cycle characteristics of the component.

Furthermore, this testbed has demonstrated the framework for routine device characterization and the functionality of a first proof-of-concept device from a standard temperature Component Manufacturer. However, further work is needed to demonstrate the process for design iteration necessary to achieve a component that meets the requirements of the System Integrator. This next phase of work will involve more detailed testing with richer analysis from the Academic Partner to inform design decisions.

Finally, while we have defined a methodology for functionally testing LNAs at 4 K, there is an evolution of TaaS that needs to take place in the form of test methodology improvements, scaling to production, reliability testing, and expanding TaaS to other cryogenic components for QC applications. Examples of such components of significance to the QC supply chain are shown in Table II and include a combination of both active and passive components in the qubit control and readout chain, as well as a list of component-relevant device specifications for testing. In each case, customized test protocols specific to each component will need to be formulated with consensus from all TaaS parties.

## VI. CONCLUSION

We demonstrated an initial TaaS workgroup model for the QC supply chain, using LNAs as a pilot test case. Our cryogenic test protocol incorporated basic qualification and device specifications reporting, including functional RF parameters and noise measurements. Some measurement errors remain; however, these will be addressed with future testing. The learnings from these measurements will be applied to future system developments for reliability and production scale testing. Overall, this workgroup demonstrated a good use case for TaaS. However, the LNA is one commodity among many that are used to build QCs. There is an immediate need for the industry to drive the creation of standardized and scalable test methodologies for components, including both active and passive cryo-components (e.g., isolators, circulators, high density interconnects and cables, etc.). We envision that over time, the industry will build a scalable supply chain and ecosystem of cryogenically qualified parts that can be used in the rapidly advancing production of QC systems.

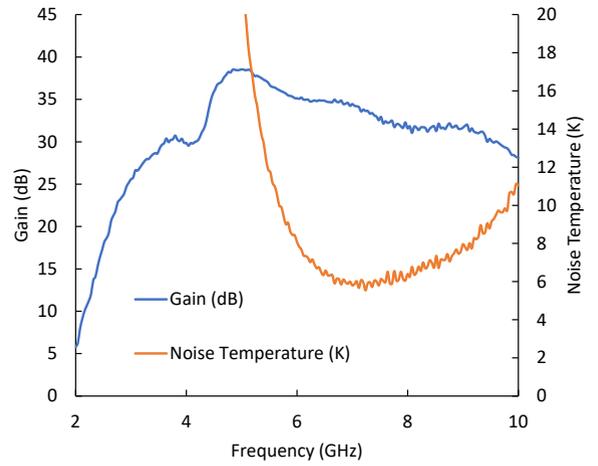

**Figure 16** Noise Temperature measurement for LNA-T. The gain curve (blue) is also extracted from the Y-factor measurement and agrees with the VNA measurements (Fig. 15). Within the 6-9 GHz operating bandwidth, the noise temperature varies between 6 K and 8 K. Our results show good agreement with the Manufacturer specifications of 6 K.

TABLE II
LIST OF COMPONENTS SIGNIFICANT TO THE QC SUPPLY CHAIN

| Component[a] | Relevant Specifications[b] |
|---|---|
| Quantum-limited amplifiers (e.g., TWPA, JPC, JPA, KIT) [40-44] | Gain / flatness Saturation power Noise figure |
| High-density wiring (CuNi or superconducting), connectors [45-47] | Loss, character. impedance Transition temperature Thermal conductivity |
| Cryogenic attenuators [48, 49] | Attenuation flatness Attenuator TCR Return loss |
| Filters [10, 50] | LP / BP / SB rejection[c] Filter frequency shift Return / insertion loss |
| Directional couplers, circulators, isolators [51, 52] | Coupler directivity / flatness Magnetic interference Return / insertion loss |
| Switches [53, 54] | Return / insertion loss Isolation, static dissipation |

[a]Where informative, developments in literature or commercial products have been provided as references.
[b]Not listed (but to be measured for all components) are standard parameters such as operating bandwidth ($BW_{op}$) and frequency range.
[c]LP: low-pass; BP: band-pass; SP: stopband.

## ACKNOWLEDGMENT

We give special thanks to Celia Merzbacher, Jon Felbinger, and other members of QED-C for organizing the technical advisory committee that spawned this workgroup. Special thanks to Niklas Wadefalk for providing the data on the control device and reviewing our analysis. We thank Del Pierson for supplying the cryogenic calibration standards. Thank you to Yves Idzerda for participating in the initial discussions that formed the workgroup model. We thank Dazhen Gu for advising on the noise figure calibration. Finally, we thank Fawad Maqbool and Shaf Khan for supplying additional samples for TME qualification.

## APPENDIX

### A. EQUIPMENT AND COMPONENTS

TABLE III
LIST OF EQUIPMENT AND COMPONENTS USED IN MEASUREMENT

| Part Number | Purpose | Equipment Vendor |
|---|---|---|
| HPD Model 106 | ADR Cryostat (4 K) | FormFactor Inc. |
| P5025B VNA | Vector Network Analzyer | Keysight Technologies |
| N9010B EXA | Signal Analyzer | Keysight Technologies |
| 346C | Calibrated Noise Source | Keysight Technologies |
| N4691B | ECal Module | Keysight Technologies |
| NC3609 | High ENR Calibrated Noise Source | Noisecom |
| R583423141 | Cryogenic SP6T Switch | Radiall USA Inc. |
| 2021-5001-33-CRYO | Cryogenic SHORT | XMA Corp. |
| XA2080-5001-02-CRYO | Cryogenic THRU | XMA Corp. |
| XA2080-5002-02-CRYO | Cryogenic LINE | XMA Corp. |
| CX-1050-CD-1.4L | Cryogenic Thermometer | Lake Shore Cryotronics Inc. |
| CX-1050-CO-HT-1.4L | Cryogenic Thermometer | Lake Shore Cryotronics Inc. |
| LNF-LNC4_8C | LNA-C (qualification) | Low Noise Factory |
| 07711 | LNA-T (spec. testing) | Marki Microwave |

### B. LIST OF ABBREVIATIONS

| | | |
|---|---|---|
| DUT | – | Device Under Test |
| ENR | – | Excess Noise Ratio |
| QC | – | Quantum Computing |
| LNA | – | Low Noise Amplifier |
| LNA-C | – | Low-Noise Amplifier (Control) |
| LNA-T | – | Low-Noise Amplifier (Test) |
| NF | – | Noise Figure |
| RF | – | Radio Frequency |
| SA | – | Spectrum Analyzer |
| TRL | – | Thru-Reflect-Line |
| TaaS | – | Test-as-a-Service |
| TME | – | Test and Measurement Equipment |
| VNA | – | Vector Network Analyzer |

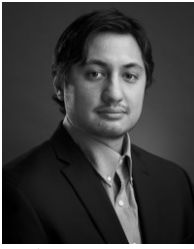
**BRANDON W. BOIKO** received his B.S./M.S. degree in Mechanical Engineering from the University of Colorado in Boulder, Colorado, USA in 2017.

From 2015 to 2020 he worked as a Cryogenic Mechanical Engineer at High Precision Devices Inc. where he designed low temperature systems used for astronomy, condensed matter physics, and high-performance computing. Since Fall of 2020, he has worked as Sr. Principal Application Engineer at FormFactor inc. in Boulder, Colorado, USA, where he is focused on developing the quantum applications for the HPD cryogenics product group. He is currently developing the Advanced Cryogenic Lab to establish cryogenic testing as a service within the quantum supply chain.

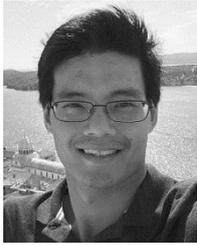
**ERIC J. ZHANG** received his B.A.Sc degree in Engineering Science (specializing in Engineering Physics) from the University of Toronto, Ontario, Canada in 2010. He received his Ph.D. in electrical engineering from Princeton University, New Jersey, USA in 2016 where he worked on infrared spectroscopy for trace-gas sensing.

From 2017 to 2019, he was a Postdoctoral Researcher at IBM T.J. Watson Research Center in Yorktown Heights, New York, USA, where his research encompassed integrated silicon photonic platforms and denoising algorithms for chip-scale sensors. Since 2019, he has remained as Research Staff Member at IBM Research, where he is actively involved in experimental quantum computing. He has co-authored over 50 journal and conference publications and holds 8 patents.

Dr. Zhang is the recipient of the Outstanding Technical Achievement Award (IBM Research) and the Walbridge Fund (Princeton University). In addition to his technical role, Dr. Zhang is the 2019 Optica Ambassador, where he provides career guidance to young professionals in photonics.

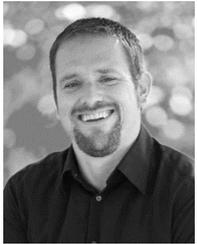
**DOUG JORGESEN** is the Vice President of Applications and System Engineering at Marki Microwave, where he has been since 2012. He has written extensively on mixers, baluns, IQ mixers, and other components. Doug received a BS in Electrical Engineering from the University of Illinois at Urbana-Champaign and PhD in Photonics from the University of California, San Diego.

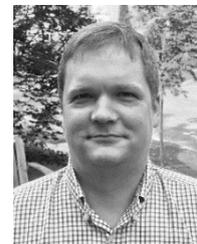
**SEBASTIAN ENGELMANN** received the Vordiplom degree in physics from the University of Würzburg, Würzburg, Germany, in 2003, and the Ph.D. degree in materials science and engineering from the University of Maryland, College Park, MD, USA, in 2008. In 2008, he joined the IBM Thomas J. Watson Research Center, Yorktown Heights, NY, USA. He joined the Advanced Plasma Processing Group, IBM T. J. Watson Research Center, where he developed plasma etch processes to support future device generations for exploratory CMOS device research and devices beyond the CMOS era. He recently joined IBM Quantum as Hardware Component Lead.

Dr. Engelmann's research activities include developing plasma processes that enable novel CMOS device geometries, novel materials (SiGe, III-V, MRAM, etc.) or disruptive, new technology approaches (Photonics packaging, neuromorphic devices, IoT applications, etc.). He is currently working on establishing a stable component supply chain for quantum computing. Dr. Engelmann is recipient of numerous Outstanding Technical Awards and Research Division awards. He is a member of SPIE and AVS. He is author on more than 90 papers and inventor on more than 35 patents.

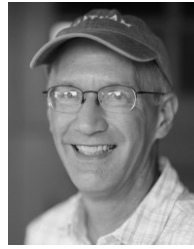
**CURTIS GROSSKOPF** received his B.A./M.A. degrees in Engineering Mechanics from the University of Wisconsin - Madison in 1986 and 1987. He joined IBM in 1988 and his career has focused on IC packaging and electronic component assembly for procured technologies and the 2nd level interaction between electronic components and printed board assembly processes. He was deeply involved in setting IBM's qualification requirements for components as the industry switched from SnPb to Pb-free soldering processes for EU RoHS and led the research used to propose the higher evaluation temperatures used by the electronics industry to qualify components for Pb-free, surface mount soldering in J-STD-020.

He is the technical lead for IBM's component Supply Chain Engineering team for environmental (RoHS, REACH, etc.) and Pb-free issues. He is currently the co-chair of IPC's 2-15f Obsolete and Discontinued Products committee and the chair of JEDEC's JC14.4 - Quality Systems committee. He recently completed leading a diverse JEDEC working group in the creation of JEDEC's JESD22-B120 Wire Bond Pull Test Methods, that was the first industry test method to propose failure criteria for Cu wirebonds, but also provide necessary guidance for the pulling of wirebond structures that had not been previously addressed in any other test method.

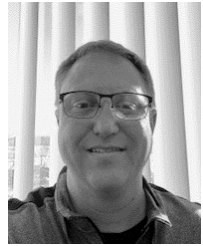
**RYAN PASKE** is a Senior Technical Staff Member in Supply Chain Engineering at IBM, where he has been since graduating college in 1998. He has performed a variety of roles at IBM in new product introduction, business transformation, and supply chain enablement. He is currently responsible for developing the IBM Supply Chain to support IBM Quantum. Ryan received a BS in Industrial Technology from the University of Wisconsin – Stout.